**Title Page**

# Brain state stability during working memory is explained by network control theory, modulated by dopamine D1/D2 receptor function, and diminished in schizophrenia

Urs Braun[1,2*] MD, Anais Harneit[1] MSc, Giulio Pergola[3] PhD, Tommaso Menara[4] MSc, Axel Schaefer[5] PhD, Richard F. Betzel[6] PhD, Zhenxiang Zang[1] MSc, Janina I. Schweiger[1] MD, Kristina Schwarz[1] MSc, Junfang Chen[1] MSc, Giuseppe Blasi[3] MD PhD, Alessandro Bertolino[3] MD PhD, Daniel Durstewitz[7] PhD, Fabio Pasqualetti[4] PhD, Emanuel Schwarz[1] PhD, Andreas Meyer-Lindenberg[1] MD, Danielle S. Bassett[2,8#] PhD, Heike Tost[1#] MD

[1] Department of Psychiatry and Psychotherapy, Central Institute of Mental Health, Medical Faculty Mannheim, University of Heidelberg, Mannheim, Germany

[2] Department of Bioengineering, University of Pennsylvania, Philadelphia, PA, USA

[3] Department of Basic Medical Science, Neuroscience, and Sense Organs, University of Bari Aldo Moro, 70124 Bari, Italy

[4] Mechanical Engineering Department, University of California at Riverside, Riverside, CA, USA

[5] Bender Institute of Neuroimaging, Justus Liebig University Giessen & Center for Mind, Brain and Behavior, University of Marburg and Justus Liebig University Giessen, Giessen, Germany

[6] Department of Psychological and Brain Sciences, Indiana University, Bloomington, IN, USA

[7] Department of Theoretical Neuroscience, Central Institute of Mental Health, Medical Faculty Mannheim/Heidelberg University, 68159 Mannheim, Germany

[8] Department of Psychiatry, Department of Neurology, Department of Physics & Astronomy, and Department of Electrical & Systems Engineering, University of Pennsylvania, Philadelphia, PA, USA

# these authors contributed equally

**Corresponding Author**: Urs Braun M.D., Department of Psychiatry and Psychotherapy, Central Institute of Mental Health, Medical Faculty Mannheim, University of Heidelberg, J5, 68159 Mannheim, Germany; Tel.: +49 621 1703 6519; e-mail: urs.braun@zi-mannheim.de

**Key Words**: network control theory, executive function, working memory, schizophrenia, dopamine

**Short title**: Control over brain states during working memory impacted by dopamine function.

**Word Count** (unreferenced Abstract): 72/70

**Word Count** (Article Body): 1488/1500

**Number of Figures**: 2/3

**Number of References**: 20/20

This paper contains Supplementary Materials.

*Dynamical brain state transitions are critical for flexible working memory but the network mechanisms are incompletely understood. Here, we show that working memory entails brain-wide switching between activity states. The stability of states relates to dopamine D1 receptor gene expression while state transitions are influenced by D2 receptor expression and pharmacological modulation. Schizophrenia patients show altered network control properties, including a more diverse energy landscape and decreased stability of working memory representations.*

Working memory is an essential part of executive cognition depending on prefrontal neurons functionally modulated through dopamine D1 and D2 receptor activation (1-3). The dual-state theory of prefrontal dopamine function links the differential activation of dopamine receptors to two discrete dynamical regimes: a D1-dominated state with a high energy barrier favoring robust maintenance of cognitive representations and a D2-dominated state with a flattened energy landscape enabling flexible switching between states (4). Recent accounts extend the idea of dopamine's impact on working memory from a local prefrontal to a brain-wide network perspective (5, 6), but the underlying neural dynamics and brain-wide interactions have remained unclear.

Network control theory (NCT) can be used to model brain network dynamics as a function of interconnecting white matter tracts and regional control energy (7). Based on the connectome, NCT can be used to examine the landscape of brain activity states: that is, which states within a dynamic scheme would the system have difficulty accessing, and more importantly, which regions need to be influenced (and to what extent) to make those states accessible (8). Specifically, to quantify accessibility, we approximate brain dynamics locally by a simple linear dynamical system, $\dot{x}(t) = \boldsymbol{A}x(t) + \boldsymbol{B}u(t)$, where $x(t)$ is the brain state inferred from functional magnetic resonance imaging (fMRI), $\boldsymbol{A}$ is a structural connectome inferred from DTI data, $u$ is the control input, and $\boldsymbol{B}$ is a matrix describing which regions enact control. To investigate, based on this conception, how the brain transitions between different cognitive states, we defined states as individual brain activity patterns related to a working memory condition (2-back) and to an attention control condition requiring motor response (0-back) in a sample of 178 healthy individuals undergoing fMRI (***Fig. S1;*** Online Methods). Further, we obtained individual structural connectomes from white matter by DTI fiber tracking, and computed the optimal control energy necessary to drive the dynamical system from the 0-back activity pattern to the 2-back pattern, or *vice versa* (***Fig. S2***).

We defined the stability of both brain states as the inverse energy necessary to revisit that state, where the energy, loosely, is defined as the average size of the control signals $u(t)$ needed to instantiate a specific trajectory in the dynamical system as defined above (see Eq. 3 & 5 in Online Methods). As expected, the cognitively more demanding 2-back state was less stable (i.e., required higher energy for maintenance) than the control state (***Fig. 1a***; repeated measures ANOVA: main effect of 0- vs. 2-back stability: $F(1,173) = 66.80$, $p < 0.001$, see Online Methods for details on all analyses). Further, the stability of the 2-back state was significantly associated with working memory accuracy (***Fig. 1b***; $b = 0.274$, $p = 0.006$), suggesting that more stable 2-back network representations support higher working memory performance. We next investigated how the brain flexibly changes its activity pattern between states. Transitioning into the cognitively more demanding 2-back state required more control energy than the opposite transition (***Fig. 1c***; repeated measures ANOVA: $F(1,174) = 27.98$, $p = 0.001$). Other analyses suggested that prefrontal and parietal cortices steer both types of transitions, while default mode areas are preferentially important for the switch to the more cognitively demanding state (***Fig. 1d***; Online Methods). These results are in line with the assumed role of frontal-parietal circuits in steering brain dynamics (9) and shifting brain connectivity patterns (10); they also emphasize the importance of the coordinated behavior of brain systems commonly displaying deactivations during demanding cognitive tasks (11).

Following from the dual-state theory of network function, the stability of task-related brain states should be related to prefrontal D1 receptor status. To estimate individual prefrontal D1 receptor expression, we utilized methods relating prefrontal cortex D1 and D2 receptor expression to genetic variation in their co-expression partner (Online Methods), thereby enabling us to predict individual dopamine receptor expression levels from genotype data across the whole genome (12, 13). We found that D1 (but not the D2) expression-related gene score predicted stability of both states (***Fig. 2a***; 0-back: $b = 0.184$, $p = 0.034$; 2-back: $b = 0.242$, $p = 0.007$, Online Methods), in line with the assumed role of D1-related signaling in maintaining stable activity patterns during task performance (4, 14).

Independent of stability, switching between different activity representations should relate to dopamine D2 receptor function. Indeed, when controlling for stability as a nuisance covariate in the regression model, the control energy of both state transitions could be predicted by the D2 (but not the D1) receptor expression gene score (***Fig. 2b***; 0- to 2-back: $b = -0.076$, $p = 0.037$; and trending for 2- to 0-back: $b = -0.134$, $p = 0.068$, Online Methods). This finding is particularly interesting, as it suggests that the function of D1 and D2 receptors are differentially, but cooperatively, involved in steering brain dynamics between different activity patterns, in line with previous research on D1 and D2 functioning in prefrontal circuits (4, 15).

Our results thus far support the notion that the brain is a dynamical system in which the stability of a state is substantially defined by cognitive effort and modulated by D1 receptor expression, while transitions between states depend primarily on D2 receptor expression. If true, such a system should be sensitive to dopaminergic manipulation, and interference with D2-related signaling should reduce the brain's ability to control its optimal trajectories, i.e. increase the control energy needed when switching between states. To test these hypotheses, we investigated an independent sample of healthy controls (n=16, ***Table S2***) receiving 400 mg Amisulpride, a selective D2 receptor antagonist, in a randomized, placebo-controlled, double-blind pharmacological fMRI study. As expected, we observed that greater control energy was needed for transitions under D2 receptor blockade (***Fig. 2c***; repeated measures ANOVA with drug and transition as within-subject factors; main effect of drug: $F(1,10) = 7.27$, $p = 0.022$; drug-by-condition interaction: $F(1,10) = 0.42$, $p = 0.665$). We observed no effect on the stability of states; that is, the inverse control energy required to stabilize a current state (main effect of drug: $F(1,8) = 0.715$, $p = 0.422$, ***Table S3***).

Dopamine dysfunction, working memory deficits, and alterations in brain network organization are hallmarks of schizophrenia (16-19). We therefore tested for differences in the state stability and in the ability to control state transitions between schizophrenia patients and a healthy control sample balanced for age, sex, performance, head motion, and premorbid IQ (see ***Table S1***). Stability in schizophrenia patients was reduced for the cognitively demanding working memory state ($F(1,98) = 6.43$, $p = 0.013$), but not for the control condition ($F(1,98) = 0.052$, $p = 0.840$, ***Table S3***). Control energy needed for the 0- to 2-back transition was significantly higher in schizophrenia (***Fig. 2d***; $F(1,98) = 5.238$, $p = 0.024$), while the opposite transition showed no significant group difference (ANOVA: $F(1,98) = 0.620$, $p = 0.433$, ***Table S3***), in line with clinical observations that D2 blockade does not ameliorate cognitive symptoms in schizophrenia (20). These results suggest that the brain energy landscape is more diverse in schizophrenia, making the system more difficult to steer appropriately. To further strengthen this notion, we estimated the variability in suboptimal (higher energy) trajectories connecting different of cognitive states (Online Methods). We expected that in a diversified energy landscape, the variation of trajectories around the minimum-energy trajectory should be larger, implying that small perturbations may have a more substantial impact. In line with our hypothesis, we found that the variability in such perturbed trajectories was indeed increased in schizophrenia (rm-ANOVA: main effect of group: $F(1,98) = 4.789$, $p = 0.031$, Online Methods).

Several aspects of our work require special consideration. Firstly, to relate brain dynamics to cognitive function, we focus on discrete brain states where each state is summarized by a single brain activation patterns rather than linear combination of multiple brain activity patterns. Secondly, although we could demonstrate a link between brain dynamics, measured by means of control energy, and predicted prefrontal dopamine receptor expression, the link is indirect and requires confirmation by direct measurements. Thirdly, we cannot exclude the possibility that disorder severity, duration, symptoms or medication may have influenced network dynamics in schizophrenia patients, although our supplemental analyses do not support this conclusion (Online Methods). Finally, while the sample sizes of our pharmacological and patient study are rather small, we were able to show comparable effects of dopaminergic manipulation on control properties using a second (Online Methods), further supporting the validity of the underlying rationale.

In summary, our data demonstrate the utility of network control theory for the non-invasive investigation of the mechanistic underpinnings of (altered) brain states and their transitions during cognition. Our data suggest that engagement of working memory involves brain-wide switching between activity states and that the steering of these network dynamics is differentially, but cooperatively, influenced by dopamine D1 and D2 receptor function. Moreover, we show that schizophrenia patients show reduced controllability and stability of working memory network dynamics, consistent with the idea of an altered functional architecture and energy landscape of cognitive brain networks.

**Acknowledgements**

The authors thank all individuals who have supported our work by participating in our studies. There was no involvement by the funding bodies at any stage of the study. We thank Oliver Grimm, Leila Haddad, Michael Schneider, Natalie Hess, Sarah Plier and Petya Vicheva for valuable research assistance. The authors thank Jason Kim and Lorenzo Caciagli for valuable feedback on the manuscript.

U.B. acknowledges grant support by the German Research Foundation (DFG, grant BR 5951/1-1). H.T. acknowledges grant support by the German Research Foundation (DFG, Collaborative Research Center SFB 1158 subproject B04, Collaborative Research Center TRR 265 subproject A04, GRK 2350 project B2, grant TO 539/3-1) and German Federal Ministry of Education and Research (BMBF, grants 01EF1803A project WP3, 01GQ1102). AML acknowledges grant support by the German Research Foundation (DFG, Collaborative Research Center SFB 1158 subproject B09, Collaborative Research Center TRR 265 subproject S02, grant ME 1591/4-1) and German Federal Ministry of Education and Research (BMBF, grants 01EF1803A, 01ZX1314G, 01GQ1003B), European Union's Seventh Framework Programme (FP7, grants 602450, 602805, 115300 and HEALTH-F2-2010-241909, Innovative Medicines Initiative Joint Undertaking (IMI, grant 115008) and Ministry of Science, Research and the Arts of the State of Baden-Wuerttemberg, Germany (MWK, grant 42-04HV.MED(16)/16/1). DSB and RBF would like to acknowledge support from the John D. and Catherine T. MacArthur Foundation, the Alfred P. Sloan Foundation, the Army Research Laboratory and the Army Research Office through contract numbers W911NF-10-2-0022 and W911NF-14-1-0679, the National Institute of Health (2-R01-DC-009209-11, 1R01HD086888-01, R01-MH107235, R01-MH107703, and R21-M MH-106799), the Office of Naval Research, and the National Science Foundation (BCS-1441502, CAREER PHY-1554488, and BCS-1631550). E.S. gratefully acknowledges grant support by the Deutsche Forschungsgemeinschaft, DFG (SCHW 1768/1-1). X.L.Z. is a Ph.D. scholarship awardee of the Chinese Scholarship Council. DD acknowledges grant support by the German Research

Foundation (DFG, Du 354/10-1). G.P. has received funding from the European Union's Horizon 2020 research and innovation program under the Marie Skłodowska-Curie No. 798181: "IdentiFication of brain deveLopmental gene co-expression netwOrks to Understand RIsk for SchizopHrenia" (FLOURISH).

The content of this paper is solely the responsibility of the authors and does not necessarily represent the official views of any of the funding agencies

## Financial disclosures

A.M.-L. has received consultant fees from Blueprint Partnership, Boehringer Ingelheim, Daimler und Benz Stiftung, Elsevier, F. Hoffmann-La Roche, ICARE Schizophrenia, K. G. Jebsen Foundation, L.E.K Consulting, Lundbeck International Foundation (LINF), R. Adamczak, Roche Pharma, Science Foundation, Synapsis Foundation – Alzheimer Research Switzerland, System Analytics, and has received lectures including travel fees from Boehringer Ingelheim, Fama Public Relations, Institut d'investigacions Biomèdiques August Pi i Sunyer (IDIBAPS), Janssen-Cilag, Klinikum Christophsbad, Göppingen, Lilly Deutschland, Luzerner Psychiatrie, LVR Klinikum Düsseldorf, LWL PsychiatrieVerbund Westfalen-Lippe, Otsuka Pharmaceuticals, Reunions i Ciencia S. L., Spanish Society of Psychiatry, Südwestrundfunk Fernsehen, Stern TV, and Vitos Klinikum Kurhessen. A.B. has received consultant fees from Biogen and speaker fees from Lundbeck, Otsuka, Recordati, and Angelini.

The remaining authors reported no biomedical financial interests of potential conflicts of interest.

## References

1. Goldman-Rakic PS (1995): Cellular basis of working memory. *Neuron*. 14:477-485.
2. Ott T, Jacob SN, Nieder A (2014): Dopamine receptors differentially enhance rule coding in primate prefrontal cortex neurons. *Neuron*. 84:1317-1328.
3. Meyer-Lindenberg A, Kohn PD, Kolachana B, Kippenhan S, McInerney-Leo A, Nussbaum R, et al. (2005): Midbrain dopamine and prefrontal function in humans: interaction and modulation by COMT genotype. *Nat Neurosci*. 8:594-596.
4. Durstewitz D, Seamans JK (2008): The dual-state theory of prefrontal cortex dopamine function with relevance to catechol-o-methyltransferase genotypes and schizophrenia. *Biol Psychiatry*. 64:739-749.
5. Arnsten AF (2011): Catecholamine influences on dorsolateral prefrontal cortical networks. *Biol Psychiatry*. 69:e89-99.
6. Roffman JL, Tanner AS, Eryilmaz H, Rodriguez-Thompson A, Silverstein NJ, Ho NF, et al. (2016): Dopamine D1 signaling organizes network dynamics underlying working memory. *Sci Adv*. 2:e1501672.
7. Kim JZ, Soffer JM, Kahn AE, Vettel JM, Pasqualetti F, Bassett DS (2018): Role of Graph Architecture in Controlling Dynamical Networks with Applications to Neural Systems. *Nat Phys*. 14:91-98.
8. Betzel RF, Gu S, Medaglia JD, Pasqualetti F, Bassett DS (2016): Optimally controlling the human connectome: the role of network topology. *Sci Rep*. 6:30770.
9. Ferenczi EA, Zalocusky KA, Liston C, Grosenick L, Warden MR, Amatya D, et al. (2016): Prefrontal cortical regulation of brainwide circuit dynamics and reward-related behavior. *Science*. 351:aac9698.
10. Cole MW, Reynolds JR, Power JD, Repovs G, Anticevic A, Braver TS (2013): Multi-task connectivity reveals flexible hubs for adaptive task control. *Nat Neurosci*. 16:1348-1355.
11. Greicius MD, Krasnow B, Reiss AL, Menon V (2003): Functional connectivity in the resting brain: a network analysis of the default mode hypothesis. *Proceedings of the National Academy of Sciences of the United States of America*. 100:253-258.
12. Fazio L, Pergola G, Papalino M, Di Carlo P, Monda A, Gelao B, et al. (2018): Transcriptomic context of DRD1 is associated with prefrontal activity and behavior during working memory. *Proc Natl Acad Sci U S A*. 115:5582-5587.
13. Pergola G, Di Carlo P, D'Ambrosio E, Gelao B, Fazio L, Papalino M, et al. (2017): DRD2 co-expression network and a related polygenic index predict imaging, behavioral and clinical phenotypes linked to schizophrenia. *Transl Psychiatry*. 7:e1006.
14. Bloemendaal M, van Schouwenburg MR, Miyakawa A, Aarts E, D'Esposito M, Cools R (2015): Dopaminergic modulation of distracter-resistance and prefrontal delay period signal. *Psychopharmacology*. 232:1061-1070.
15. Trantham-Davidson H, Neely LC, Lavin A, Seamans JK (2004): Mechanisms underlying differential D1 versus D2 dopamine receptor regulation of inhibition in prefrontal cortex. *J Neurosci*. 24:10652-10659.
16. Howes OD, Kapur S (2009): The dopamine hypothesis of schizophrenia: version III--the final common pathway. *Schizophrenia bulletin*. 35:549-562.
17. Barch DM, Smith E (2008): The cognitive neuroscience of working memory: relevance to CNTRICS and schizophrenia. *Biol Psychiatry*. 64:11-17.
18. Tost H, Alam T, Meyer-Lindenberg A (2010): Dopamine and psychosis: theory, pathomechanisms and intermediate phenotypes. *Neurosci Biobehav Rev*. 34:689-700.
19. Braun U, Schafer A, Bassett DS, Rausch F, Schweiger JI, Bilek E, et al. (2016): Dynamic brain network reconfiguration as a potential schizophrenia genetic risk mechanism modulated by NMDA receptor function. *Proc Natl Acad Sci U S A*. 113:12568-12573.
20. Millan MJ, Fone K, Steckler T, Horan WP (2014): Negative symptoms of schizophrenia: clinical characteristics, pathophysiological substrates, experimental models and prospects for improved treatment. *European neuropsychopharmacology : the journal of the European College of Neuropsychopharmacology*. 24:645-692.

## Figures

**Figure 1:**

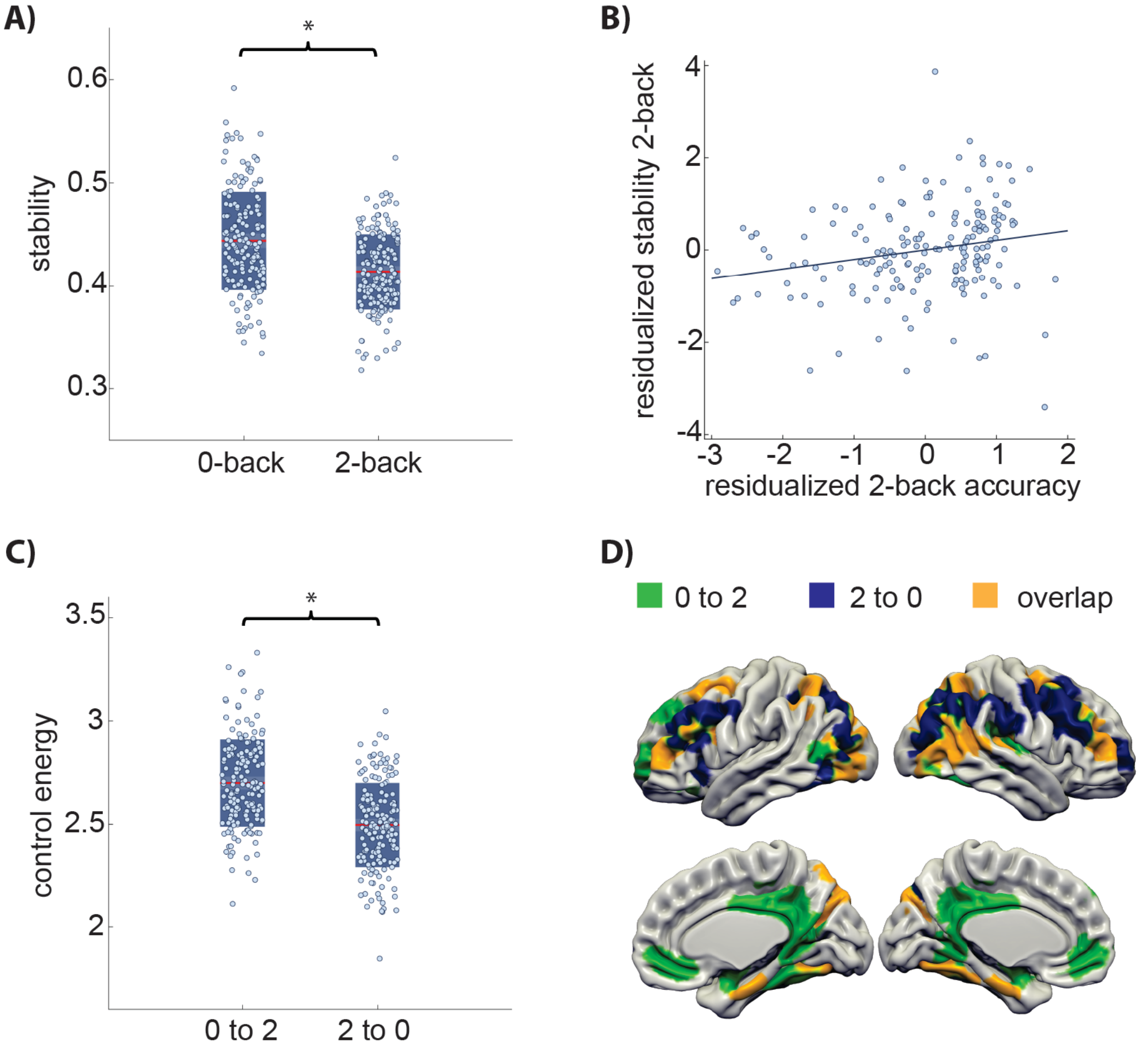

**Controllability and stability of brain dynamics during working memory**

A) The stability of the 2-back state reflecting working memory activity is lower than that of the 0-back state reflecting motor and basic attention control activity ($F(1,173) = 66.80$, $p < 0.001$). Red lines indicate mean values and boxes indicate one standard deviation of the mean. B) Associations of 2-back stability with working memory performance (accuracy: $b = 0.274$, $p = 0.006$; covarying for age, sex, and mean activity). C) Steering brain dynamics from the control condition to the working memory condition requires more control energy than *vice versa* ($F(1,174) = 27.98$, $p < 0.001$). D) Unique and common sets of brain regions contribute most to the transition from 0-back to 2-back and the transition from 2-back to 0-back transitions, respectively. For illustrative purposes, we projected the computed control impact of each brain region (Online Methods) for the respective transitions on a 3D structural template, displaying the 20% highest for each transition.

**Figure 2:**

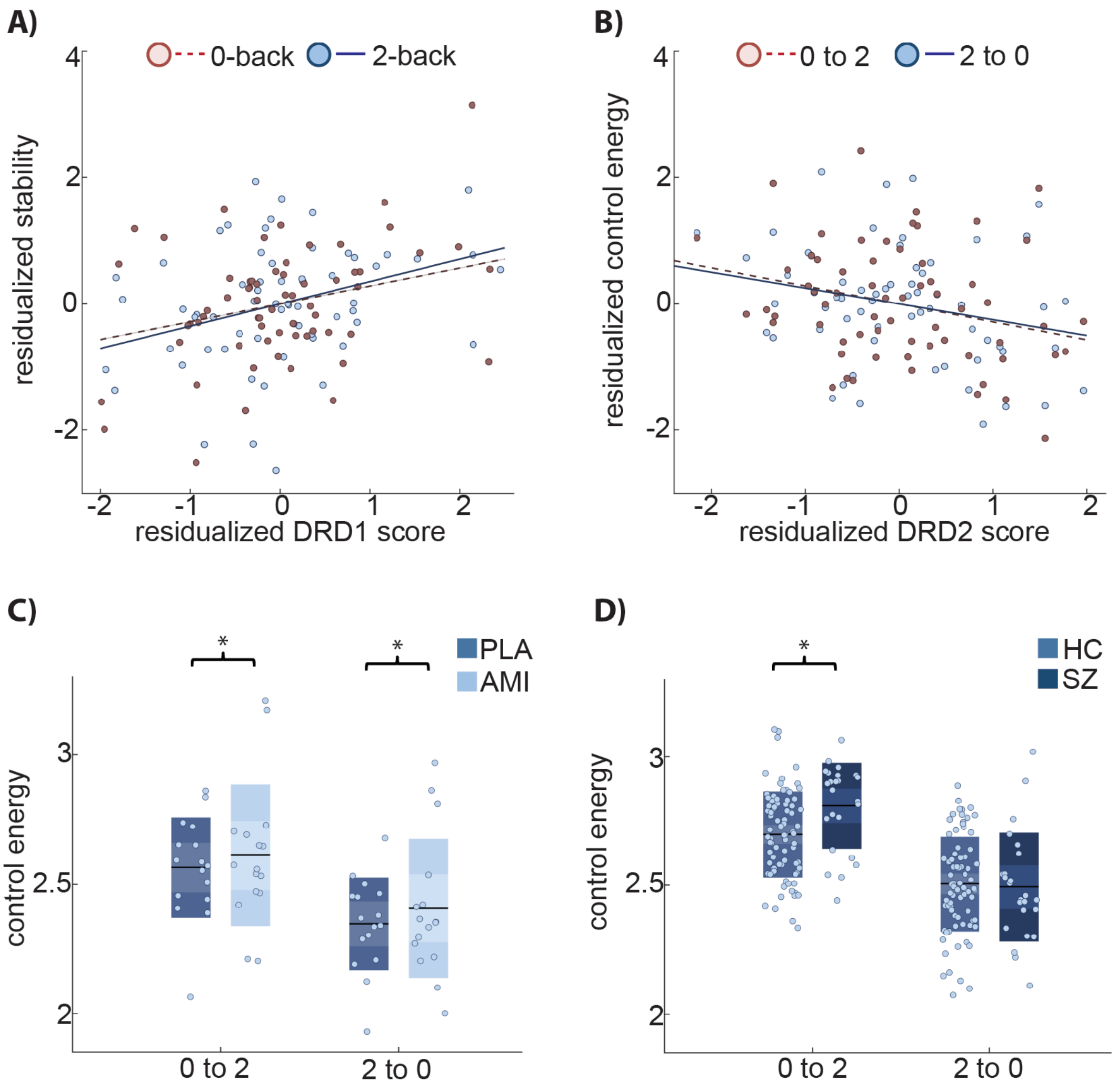

**Dopamine receptor expression and pharmacological modulation impact whole brain dynamics**

A) Genetic scores predicting DRD1 expression in prefrontal regions positively predict stability of both brain states (0-back: *b* = 0.184, p = 0.034; 2-back: *b* = 0.242 p = 0.007; age, sex, mean brain state activity, first 5 genetic PCA components as covariates of non-interest). B) Genetic scores predicting DRD2 expression in prefrontal regions negatively predict control energy for both brain state transitions (0-back to 2-back: *b* = -0.076, p = 0.037; and trend wise for 2-back to 0-back: *b* = -0.134, p = 0.068; age, sex, mean brain activity difference, first 5 genetic PCA components, stability of 0-back and 2-back as covariates of non-interest). C) Amisulpride increases control energy for transitions in comparison to placebo (main effect of drug: $F(1,10) = 7.27$, $p = 0.022$; interaction drug by condition: $F(1,10) = 0.42$, $p = 0.665$, activity difference, drug order, and sex as covariates of non-interest). Black lines indicate mean values and boxes indicate one standard deviation of the mean. D) Schizophrenia patients need more control energy when transitioning into the working memory condition than matched healthy controls ($F(1,98) = 5.238$, $p = 0.024$, age, sex, tSNR and mean activity as covariates of non-interest), but not *vice versa*.

# ONLINE METHODS

*Urs Braun et al. “Brain state stability during working memory is explained by network control theory, modulated by dopamine D1/D2 receptor function, and diminished in schizophrenia“*

## 1. Participants and study design

All participants provided written informed consent for protocols approved by the Institutional Review Board of the medical faculty in Mannheim. For the first study including healthy controls and patients with schizophrenia, a total of 202 subjects (178 healthy controls, 24 schizophrenia patients) were included (see Table S1). General exclusion criteria in controls included the presence of a lifetime history of psychiatric, neurological, or significant general medical illness, pregnancy, a history of head trauma, and current alcohol or drug abuse. The patients were recruited from the Department of Psychiatry and Psychotherapy at the Central Institute of Mental Health in Mannheim and via local advertisements. A trained psychiatrist or psychologist verified the diagnosis of schizophrenia based on ICD-10 criteria.

Neuropsychological characterization of healthy controls included the trail-making-test B (TMT-B) (1) and the German multiple-choice vocabulary intelligence test (MWT-B) (2) as a measure of premorbid IQ. Clinical characterization included the assessment of current symptom severity using the Positive and Negative Symptom Scale (PANSS) (3), Beck's Depression Inventory (4), measures of global functioning in daily life (global assessment of function (GAF)) and current antipsychotic medication dosage (converted into chlorpromazine dose equivalents (CPZE)).

For the second, pharmacological intervention study, 17 healthy individuals completed a subject- and observer-blind, placebo-controlled, randomized three-period cross-over study (see Table S2). Exclusion criteria included a regular consumption of drugs or history of drug or alcohol abuse; systolic blood pressure (SBP) greater than 140 or less than 90 mm Hg, and diastolic blood pressure (DBP) greater than 90 or less than 50 mm Hg; notable resting bradycardia (heart rate (HR) <40 bpm) or tachycardia (HR >90 bpm); use of any medication or herbal remedies taken within 14 days prior to randomization into the study or 5 times the elimination half-life of the medication, clinically significant abnormalities in laboratory test results (including hepatic and renal panels, complete blood count, chemistry panel and urinalysis): a history or presence of clinically significant ECG abnormalities (e.g. PQ/PR interval >210 ms, QTcF >450 ms) or cardiovascular disease (e.g. cardiac insufficiency, coronary artery disease, cardiomyopathy, hypokalemia, congestive heart failure, family history of congenital long QT syndrome, family history of sudden death); any personal or familial history of seizures, epilepsy or other convulsive condition, previous significant head trauma, or other factors predisposing to seizures; disorders of the central nervous system, cerebrovascular events, Parkinson's disease, migraine, depression, bipolar disorder, anxiety, any other psychiatric disorders or behavioral disturbances; regular smoking (>5 cigarettes, >3 pipe-fulls, >3 cigars per day); habitual caffeine consumption of more than 400 mg/d (approximately 4 cups of coffee or equivalent); a history or evidence of any clinically significant endocrinological, hepatic, renal, autoimmune, pulmonary, gastrointestinal, urogenital, oncological, hematological or any other disease; or a body mass index (BMI) of over 30 or below 22.

Participants were invited for a fixed interval of 7 days with each scanning session taking place at approximately the same time of day. On each of three scanning visits, individuals either received a single oral dose of 400 mg Amisulpride, 3 mg Risperidone or Placebo. MRI scanning took place 2 hours after drug administration, with the N-back paradigm commencing approximately 10 min after the start of the scan. One subject was excluded from the analysis due to an excessive body-mass index (BMI > 30).

## 2. Data acquisition

### 2.1. fMRI

For the first study, BOLD fMRI was performed on a 3T Siemens Trio (Erlangen, Germany) in Mannheim, Germany. Prior to the acquisition of functional images, a high-resolution T1-weighted 3D MRI sequence was conducted (MPRAGE, slice thickness=1.0 mm, FoV = 256 mm, TR = 1570ms, TE = 2.75 ms, TI = 800ms, $\alpha$ = 15°). Subsequently, functional data was acquired during performance of the N-back paradigm using an echo-planar imaging (EPI) sequence with the following scanning parameters: TR/ TE = 2000/30 ms, $\alpha$ =80°, 28 axial slices (slice-thickness = 4 mm + 1 mm gap), descending acquisition, FoV = 192 mm, acquisition matrix = 64 x 64 128 volumes.

For the second study, BOLD fMRI was performed on a 3T Siemens Trio Scanner (Erlangen, Germany) using an echo-planar imaging (EPI) sequence with the following parameters: TR = 1790 ms, TE = 28 ms, 34 axial slices per volume, voxel size = 3 x 3 x 3 mm, 1 mm gap, 192 x 192 mm field of view, 76° flip angle, descending acquisition.

The visual N-back paradigm is a well-established and reliable working memory task consisting of a high memory load (2-back) and an attention control condition requiring motor response (0-back) (5-7). Specifically, a diamond-shaped stimulus containing a number from 1 to 4 was presented every 2 seconds (see Figure S1). In the 0-back condition, subjects were required to press the button on the response box corresponding to the number currently displayed on the presentation screen. In the 2-back condition, subjects were required to press the button on the response box corresponding to the number presented two stimuli before the number currently displayed on the presentation screen. Stimuli were presented in alternating blocks of either 0-back or 2-back conditions. In each condition block, 14 stimuli were presented. Each condition block was repeated 4 times. Task performance was measured by accuracy (defined as the percent of correct answers) and reaction time (defined as the time span between stimulus onset and button press) for each condition separately.

### 2.2. Diffusion Tensor Imaging (DTI)

DTI data were acquired by using spin echo EPI sequences with the following parameters: TR 14000ms, TE 86ms, 2mm slice thickness, 60 non-collinear directions, b-value 1000 s/mm2, 1 b0 image, FOV 256 mm.

## 3. Atlas construction

To combine structural and functional brain imaging data, we first constructed a brain atlas that equally well respects functional and anatomical features. We transformed a recently published multimodal atlas (8) into a volumetric format by projecting its FreeSurfer pial cortex coordinates into standard MNI space. A grey matter prior probability map (thresholded at 0.3) as provided by SPM was used to define relevant voxels. Voxels were labeled by choosing the closest label with maximum distance of 4 mm. Since the published multimodal atlas does not cover all subcortical regions of interest (e.g. amygdala, thalamus), we complemented it with subcortical structures from the Harvard-Oxford atlas as implemented in FSL (9). Combining the two atlases resulted in 374 regions that covered cortical and subcortical structures. A full list of regions included in the combined atlas can be found in Supplemental Table S3.

## 4. Connectome construction

For the DTI data, the following preprocessing steps were performed with standard routines implemented in the software package FSL (9): i) correction of the diffusion images for head motion and eddy currents by affine registration to a reference (b0) image, ii) extraction of non-brain tissues (10), and iii) linear diffusion tensor fitting. After estimation of the diffusion tensor, we performed deterministic whole-brain fiber tracking as implemented in DSI Studio using a modified FACT algorithm (11). For each subject, 1,000,000 streamlines were initiated. Streamlines with a length of less than 10 mm were removed (12). For the construction of structural connectivity matrices, the brain was parcellated into 374 regions. To map these parcellations into subject space, we applied a nonlinear registration implemented in DSI Studio. We estimated the structural connectivity between any two regions of the atlas by using the mean fractional anisotropy values between respective brain regions. This procedure resulted in a weighted adjacency matrix $\boldsymbol{A}$ whose entries $\boldsymbol{A}_{ij}$ reflect the strength of structural connectivity between two brain regions.

## 5. Brain state definition

Because we were interested in investigating how the brain controls and transitions between global brain states underlying circumscribed cognitive processes (such as those supporting working memory, attention, and motor behavior), we defined brain states as stationary patterns of activity during execution of these processes. It is important to note that the temporal resolution of fMRI and the design of the N-back task limits the investigation of differential cognitive processes contributing to memory performance on fast time scales. Therefore, we cannot investigate the detailed temporal dynamics of working memory processes. However, our simplified design allows us to extract meaningful brain states elicited by a controlled cognitive process and therefore enables us to relate brain dynamics to cognitive function.

Specifically, we defined individual brain states as spatial patterns of beta estimates associated with activity across brain regions of interest during both conditions of the N-back task (13). For that purpose, after standard preprocessing procedures in SPM12/8 (including realignment to the mean image, slice time correction, spatial normalization into standard stereotactic space, resampling to 3 mm isotropic voxels, and smoothing with an 8 mm full-width at half-maximum Gaussian Kernel) (5, 6, 14), we estimated standard first-level general linear models for the N-back task, separately for each individual. Except for the SPM version, both studies followed the same preprocessing procedure. These GLM models included regressors for the 0-back and 2-back conditions of interest, as well as the 6 motion parameters as regressors of non-interest. To define the brain activity pattern associated with each condition of the task, we extracted GLM (beta) parameter estimates for the 0-back and 2-back conditions separately (13) and we averaged them across all voxels in each of the 374 regions without applying any threshold. This procedure yielded a 374 by 1 vector for each condition ($x_{0\text{-back}}$, $x_{2\text{-back}}$) per subject representing how strongly the BOLD response in each brain region was associated with the working memory (2-back) or the motor control condition (0-back), respectively. These vectors ($x_{0\text{-back}}$, $x_{2\text{-back}}$) defined the final and target brain states for the following network control analyses.

## 6. Network control theory

### 6.1. Optimal control theory framework

To model the transition between 0-back and 2-back brain states, we used the framework of optimal control, following prior work (15-17) implemented in MATLAB. Based on individual brain states $X=[x_1,...x_n]$ (in our case simplified to $n$ = 2 states: 0-back and 2-back, see above) and a structural brain network $\boldsymbol{A}$ for each subject, we approximated the local brain dynamics by a linear continuous-time equation:

$$(1)\quad \dot{x}(t) = \boldsymbol{A}\, x(t) + \boldsymbol{B}\, u(t)$$

to model the flow among task-related brain activity states. In the model, $x(t)$ is the state of the system at time $t$, $\boldsymbol{A}$ is the wiring diagram of the underlying network, $\boldsymbol{B}$ denotes an input matrix defining the control nodes, and $u(t)$ is the time-dependent control signal (17, 18). Note that while the initial state ($x_0$) and the target state ($x_T$) are empirically defined, any states of the system at other times are virtual intermediate steps in the trajectories of the state-space model. The problem of finding an optimal control energy $u^*$ that induces a trajectory from an initial state $x_0$ to a target state $x_T$ reduces to the problem of finding an optimal solution to the minimization problem of the corresponding Hamiltonian:

$$(2)\ \min[\mathrm{H}(p, x, u, t) = x^T x + \tilde{n} u^T u + p^T (\mathbf{A}x + \mathbf{B}u)]\ (15).$$

For simplicity, we set the input matrix $\boldsymbol{B} = \boldsymbol{I}_{N \times N}$, the identity matrix, allowing all brain regions to be independent controllers (16, 18).

### 6.2. Control energy, stability, and impact

Control energy for each node $k_i$, $i = 1 \ldots m$ (m = total number of brain nodes), was defined as

$$(3) \quad E_{k_i} = \int_{t=0}^{T} \left\| u^*{}_{k_i}(t) \right\|^2 dt,$$

i.e. the squared integral over time of energy input that the node has to exhibit to facilitate the transitions from the initial state to the target state (15, 16, 18). While the neurobiological foundations of control properties in the brain are not yet well understood, control energy can be interpreted as the effort of a brain region needed to steer the activity pattern of itself and its connected brain regions into the desired final activation state, for example by tuning their internal firing or activity patterns by recurrent inhibitory connections. Accordingly, the total control energy for the entire brain was defined as the sum of all control energy across all nodes

$$(4) \; E = \sum_{i=1}^{m} E_{k_i},$$

yielding one value for each transition per subject. To ensure a normal distribution of metric values for subsequent statistical testing, we applied a logarithmic transformation (base 10) to the control energy(16).

From the control energy, we can also obtain the control stability and control impact. Stability was defined as

$$(5) \quad S = \frac{1}{log_{10}(E_{x_o = x_T})},$$

i.e. the inverse control energy needed to *maintain* a state, or in other words, the control energy needed to go, e.g., from 2-back to 2-back (18). To investigate the influence that a single brain region has on the entire system's dynamics during state trajectories, we computed the control impact of each node by iteratively removing one brain region from the network and re-computing the change in control energy (15).

## 6.3. Suboptimal trajectories

To investigate the energy landscape surrounding the minimum energy (in this sense 'optimal') trajectories, we quantified the variability of suboptimal trajectories by adding subtle random perturbation to the minimum energy trajectories over 200 iterations. Note that we employed a discrete-time dynamical system rather than a continuous-time system for these analyses, as discrete-time systems are computationally more tractable. To discretize our linear continuous-time system, we employed the following transformations

(6) $\bar{\boldsymbol{A}} \triangleq e^{\boldsymbol{A}T_s}$ and

(7) $\bar{\boldsymbol{B}} \triangleq \left(\int_0^{T_s} e^{A(T_s-\tau)}\, d\tau\right)\boldsymbol{B}$,

where $\bar{\boldsymbol{A}}$ and $\bar{\boldsymbol{B}}$ are the corresponding structural matrix and control input matrix in a discrete time system and $T_s$ is the sampling time. As there is no prescriptive way to choose $T_s$, we estimated $T_s = \frac{1}{10} RT$, where $RT$ is the rise time of its fastest mode, i.e. the time that the system requires to go from 10% to 90% of its fastest step response. In the resulting discrete-time dynamical system

(8) $\bar{x}\,(t+1) = \bar{\boldsymbol{A}}\bar{x}(t) + \bar{B}\bar{u}(t)$,

at each discrete time step *t*, we applied a principal component analysis (PCA) to the cloud of suboptimal points to reduce dimensionality. Because the previous simulations show that the first PCA component explains more than 90% of variance, we continued by using the first component as a summary measure of suboptimal trajectories. To estimate the relative distance between suboptimal and optimal trajectories, we computed the percentage of variation of the maximum distance of the projected suboptimal points on the first principal component from the optimal one. This procedure resulted in a normalized measure giving the percentage of deviations for the 200 suboptimal trajectories from the optimal trajectory. Due to the heuristic nature of the algorithm applying a random perturbation, results can vary from run to run. Therefore, to increase replicability of our results, we repeated our analysis 10 times per subject (10*200 = 2,000 suboptimal trajectories per subject).

Because DTI data was not acquired during the pharmacological intervention study, we used the average connectivity matrix across all healthy subjects from study 1 to model the transition between individual brain states in the pharmacological intervention study.

## 7. Gene based polygenic co-expression indices

### 7.1. Genotyping, imputation and quality control

In this study, we used human GWAS data of 63 healthy subjects who were genotyped using HumanHap 610 and 660w Quad BeadChips. For all subjects, standard quality control (QC) and imputation were performed using the Gimpute pipeline (Chen, Lippold *et al.* 2018) and the following established QC steps were applied. **Step 1:** Determine the number of male subjects for each heterozygous SNP of the chromosome 23 and remove SNPs whose number is larger than 5% of the number of male samples. **Step 2:** Determine the number of heterozygous SNPs on X chromosome for each male sample, and remove samples that have the number of heterozygous SNPs larger than 10. **Step 3:** Remove SNPs with missing genotyping rate > 5% before sample removal. **Step 4:** Exclude samples with missingness >= 0.02. **Step 5:** Exclude samples with autosomal heterozygosity deviation |Fhet| >= 0.2. **Step 6:** Remove SNPs with the proportion of missing genotyping > 2% after sample removal. **Step 7:** Remove SNPs if the Hardy-Weinberg equilibrium exact test P-value was $< 1 \times 10^{-6}$. **Step 8:** Principal component analysis (PCA) was applied to detect population outliers. Imputation was carried out using IMPUTE2/SHAPEIT (19-21), with a European reference panel for each study sample in each 3 Mb segment of the genome. This imputation reference set is from the full 1000 Genome Project dataset (August 2012, 30,069,288 variants, release “v3.macGT1”). The length of the buffer region is set to be 500 kb on either side of each segment. All other parameters were set to default values implemented in IMPUTE2. After imputation, SNPs with high imputation quality (INFO >= 0.6) and successfully imputed in >= 20 samples were retained. From the final well-imputed dataset with 63 subjects, we extracted 8 SNPs for DRD2 and 13 SNPs for DRD1 (22-24). For the subsequent genetic imaging analyses, we only used subjects for whom both data modalities (DTI and fMRI data) were available.

### *7.2.* Polygenic co-expression index calculation

Previous publications have shown that gene sets defined using co-expression networks and selected for their association with the genes *DRD1* and *DRD2* provided replicable predictions of n-back-related brain activity and behavioral indices (23, 25-27). Weighted Gene Co-expression Network Analysis [WGCNA (28)] applied on the Braincloud dataset (N=199) of post-mortem DLPFC gene expression (29) identified 67 non-overlapping sets of genes based on their expression pattern. The co-expression gene sets including *DRD1* and *DRD2* were summarized into Polygenic Co-expression Indices (PCIs) based on SNPs that predicted co-expression of these genes (called co-expression quantitative trait loci, or co-eQTLs).

PCIs are a proxy for the assessment of the genetic component of gene transcription co-regulation and are computed as a weighted average of the effect of all genotypes of an individual among those selected in the data mining study as co-eQTLs. The effect of individual SNPs is computed as the difference between the gene co-expression distribution of minor allele carriers (heterozygotes and homozygotes) and that of major allele homozygotes, using common tools from signal detection theory (30). Genotype weights, therefore, represent the deviation in gene co-expression from a reference distribution and are not constrained by allele dose. For each genotype of each SNP we computed an index, called A', proportional to the expression of the gene of interest (*DRD1* or *DRD2*) within its co-expression module. The A' index is less dependent than d' on the assumption of a normal distribution of gene expression in each genotypic population (25). Both PCI-based predictions were significantly replicated in an independent post-mortem dataset, while controlling for ethnicity. The translational effect of these two scores on brain activity during n-back has been assessed and replicated across multiple samples, which combined amount to approximately 600 participants (22-24).

It is important to note that these dopamine-related genetic effects are large in magnitude compared to those estimated by polygenic risk score approaches that focus on epidemiological data, rather than on molecular processes. The *DRD2*-PCI we developed (23) yielded an effect size f = 0.30 in our n-back discovery sample (required sample size to obtain 80% power with α = 0.05 and covariates as in the current work: N = 71). Results were replicated in an independent fMRI dataset collected at a different institution with f = 0.20 (required sample size computed as above: N = 156). Our follow-up work on the *DRD2*-PCI (26) considered two datasets of 50 individuals each and yielded a minimum effect size f = 0.28 (in the replication sample; required sample size computed as above: N = 81). The *DRD1*-PCI was also tested in two independent samples (25), yielding a minimum effect size f = 0.37 (in the replication sample; required sample size computed as above: N = 46). Taken together, these published results show that the effects of these polygenic indices on n-back activity in the prefrontal cortex are relatively large, with sizes ranging between 0.20 and 0.37 and with required samples ranging from 46 to 156 individuals. Importantly, the *DRD2*-PCI was also tested in a small sample of 29 patients with SCZ and yielded results consistent with the effects discovered in healthy controls (23). Although the required sample sizes were computed based on the top cluster, it should be borne in mind that the technique we used in this work employs the entire brain, and therefore (i) is not subject to correction for multiple comparisons, as reflected in the uncorrected alpha used for the power calculations and (ii) benefits from the greatest possible amount of information about brain states.

## 8. Statistical inference

Statistical inference was performed using the Statistical Package for the Social Sciences 24 (IBM SPSS Statistics for Windows, Version 24). All statistical comparisons were performed while controlling for age and sex and were tested two-sided. Because we were interested in control properties of brain state transitions independent of differences in the amount of mere activation, we controlled for the respective parameters reflecting the individual differences in activations in all analyses involving control properties. In particular, for all analyses involving stability measures, we additionally controlled for the average brain energy defined as the GLM parameter estimates over all regions in the 0-back and 2-back conditions. As the control energy of a transition depends highly on the absolute difference in energy between its initial state and its final state (i.e. in our case the brain-wide activation difference between both task conditions) and as we were interested in the unique control properties independent of traditional activation differences, we additionally controlled for the difference in the mean baseline energy in each analysis involving control energy. Baseline energy was defined as the absolute average difference in the unthresholded GLM parameter estimates over all regions between 0-back and 2-back conditions. As the same arguments apply to stability measures, we controlled for the respective baseline energy of each condition in each analysis involving stability.

In all analyses involving polygenic scores for D1-/D2-expression, we further used the first 5 principal components from the PCA on the linkage-disequilibrium pruned set of autosomal SNPs to control for population stratification. Differences between control energy/stability of both conditions were assessed using repeated-measures ANOVA with condition as a within-subject factor. Drug effects were modeled using a repeated measures ANOVA with drug and condition as within-subject factors. As healthy control and schizophrenia patients differed in one of the quantified DTI imaging quality parameters (tSNR, see Table S1), we also controlled for tSNR in each analysis involving both groups.

# 9. Control analyses

## 9.1. Null models of structural brain networks

To study the impact of structural brain networks on control properties, we repeated the computation of control energy using a randomized null model of the individuals' structural brain networks that preserves the average weight, the strength distribution and the degree distribution. Null models were created using the null_model_und_sign function as implemented in the Brain Connectivity Toolbox (https://sites.google.com/site/bctnet/). In line with our expectation, control energy increased significantly for randomized networks in both groups (repeated measures ANOVA with null_model_vs_data and transition as within-subject factors, HC: main effect of null_model_vs_data, $F(1,174) = 38.284$, $p < 0.001$; SZ: $F(1,20) = 3.561$, $p = 0.074$), suggesting that human brain structural networks are in some form optimized to control brain state transitions, but schizophrenia patients tend to have less optimal, more random structural brain networks.

## 9.2. Null models of spatial activity patterns

To study the impact of the spatial distribution of activity patterns on control properties, we repeated the computation of control energy and spatially randomized individuals' brain activation patterns. In line with our expectation, control energy increased significantly for randomized networks in both groups (repeated measures ANOVA with model_vs_data and transition as within-subject factors, HC: main effect of model_vs_data, $F(1,174) = 6.995$, $p = 0.009$; SZ: $F(1,20) = 0.019$, $p < 0.0893$), suggesting that the spatial distribution of brain activity patterns is important for minimizing control effort.

## 9.3. Robustness to choice of parcellation scheme

To demonstrate the robustness of the results to our choice of parcellation scheme, we repeated our analysis using a recently published functionally defined atlas comprising a similar number of areas (31). Specifically, we used the "Gordon" template (31) consisting of 333 regions that are functionally derived from resting-state connectivity analyses. Data were reprocessed using the same pipeline as for the main analysis and all parameters were kept identical in the subsequent analysis. Notably, we replicated all main results (see Table S4), indicating that our reported findings are robust to the choice of parcellation scheme.

### 9.4. Robustness to choice of edge definition

To demonstrate the robustness of our results to our selection of connectivity measure, we repeated our analysis using the number of streamlines normalized by the respective size of the regions to construct structural connectivity matrices (15). All parameters were kept identical in the subsequent analysis. All main results could be replicated (see Table S4), indicating that our findings are robust to the choice of edge definition.

### 9.5. Impact of medication and duration of illness on control properties

In patients, the potential relationship between control energy and stability, antipsychotic drug dose (expressed in chlorpromazine equivalents (CPZE), n=20), and clinical parameters (illness duration, illness severity as indexed by global functioning (GAF) and Positive and Negative Symptom Scale (*PANSS*)) were explored using Pearson correlation. Neither the control energy for the 0-back to 2-back transition nor the opposite transition or the stability of either state were significantly associated with CPZE (N = 20, 0- to 2-back: r = 0.078, p = 0.767; 2- to 0-back: r = 0.320, p = 0.210; 0- back stability: r = 0.150, p = 0.564; 2- back stability: r = 0.096, p = 0.713), with illness duration (N = 23, 0- to 2-back: r = 0.017, p = 0.937; 2- to 0-back: r = -0.226, p = 0.299; 0- back stability: r = 0.110, p = 0.644; 2- back stability: r = 0.281, p = 0.230), or with GAF (N = 24, 0- to 2-back: r = -0.086, p = 0.690; 2- to 0-back: r = -0.254, p = 0.230; 0- back stability: r = -0.135, p = 0.570; 2- back stability: r = 0.066, p = 0.793).

### 9.6. Pharmacological validation using Risperidone

To demonstrate the robustness of our pharmacological intervention of dopaminergic signaling, we additionally analyzed the data of the Risperidone condition in the same subjects. Risperidone also preferentially targets D2 receptors, but also affects D1, adrenergic, serotoninergic and histaminergic pathways. Using the same models and covariates as in the main analysis, we detected a trend-wise increase in control energy needed for both transitions (repeated measures ANOVA with drug and transition as within-person factors; main effect of drug: F(1,10) = 3.490, p = 0.091; drug-by-condition interaction:

F(1,10) = 0.238, p = 0.636; activity difference, drug order, and sex as covariates of no interest), but no effect on stability (F(1,8) = 0.105, p = 0.334; mean brain activity, sex, and drug order as covariates of no interest). Although these results showed only trend-wise significance, likely due to the lower D2-specificity of Risperidone, the detected pattern was conserved across drugs, validating the proposed underlying concepts.

### 9.7. Null results for gene score and imaging associations

As mentioned in the main text, D1 receptor expression-related gene scores predicted stability of both states (0-back: b = 0.184, p = 0.034; 2-back: b = 0.242 p = 0.007), but not D2 receptor expression-related gene scores (0-back: b = 0.153, p = 0.109; 2-back: b = -0.01 p = 0.924). In turn, the control energy of both state transitions could be predicted by the D2 receptor expression-related score (0- to 2-back: b = -0.076, p = 0.037; and trending for 2- to 0-back: b = -0.134, p = 0.068), but not by the D1 receptor expression-related gene score (0- to 2-back: b = -0.037, p = 0.324; 2- to 0-back: b = -0.06, p = 0.418).

### 9.8. Suboptimal trajectories

As mentioned in the main text, the variability in suboptimal trajectories was greater in schizophrenia (rm-ANOVA: main effect of group: F(1,98) = 4.789, p = 0.031, controlling for age, sex, DTI tSNR ,brain state energy difference). These results remained significant after additionally accounting for the stability of both states and for the control energy of both transitions (rm-ANOVA: main effect of group: F(1,95) = 11.2, p = 0.001).

# 10. Supplementary figures

## 10.1. Figure S1: N-back task design

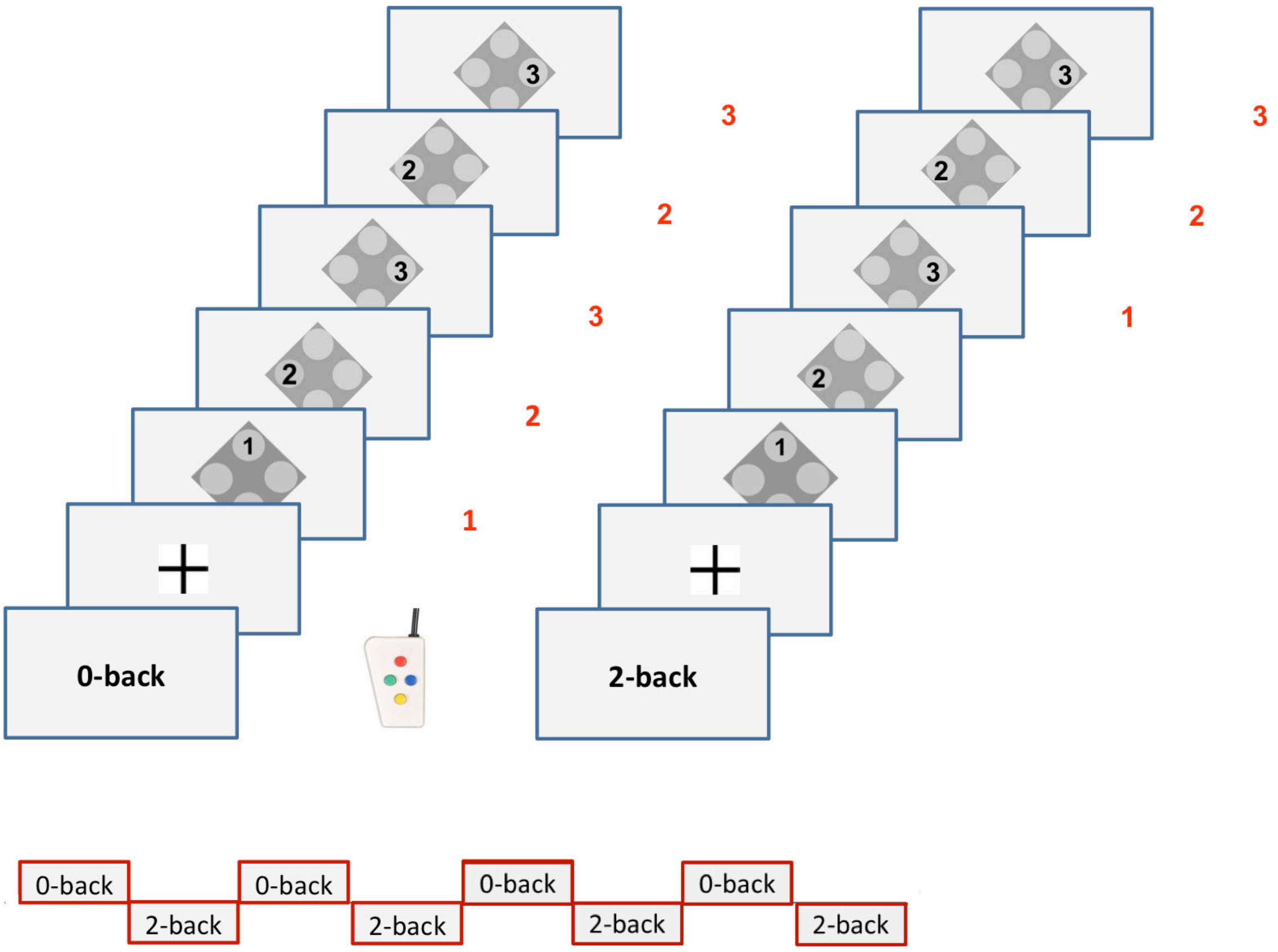

Design of the N-back task: Stimuli were presented in blocks of either 0-back (left) or 2-back (right) conditions. There was no additional control or resting condition. In the 0-back condition, subjects were instructed to press the button on the response box corresponding to the number currently displayed on the presentation screen. Here, the red numbers next to the screen images on the left indicate correct responses. In the 2-back condition, subjects were instructed to press the button on the response box corresponding to the number presented two stimuli before the number currently displayed on the presentation screen. Here, the red numbers next to the screen images on the right indicate the correct responses. Each condition block lasted 30 seconds and was repeated four times in an interleaved manner as shown on the bottom of the figure.

### 10.2. Figure S2: Network control theory concept and methods

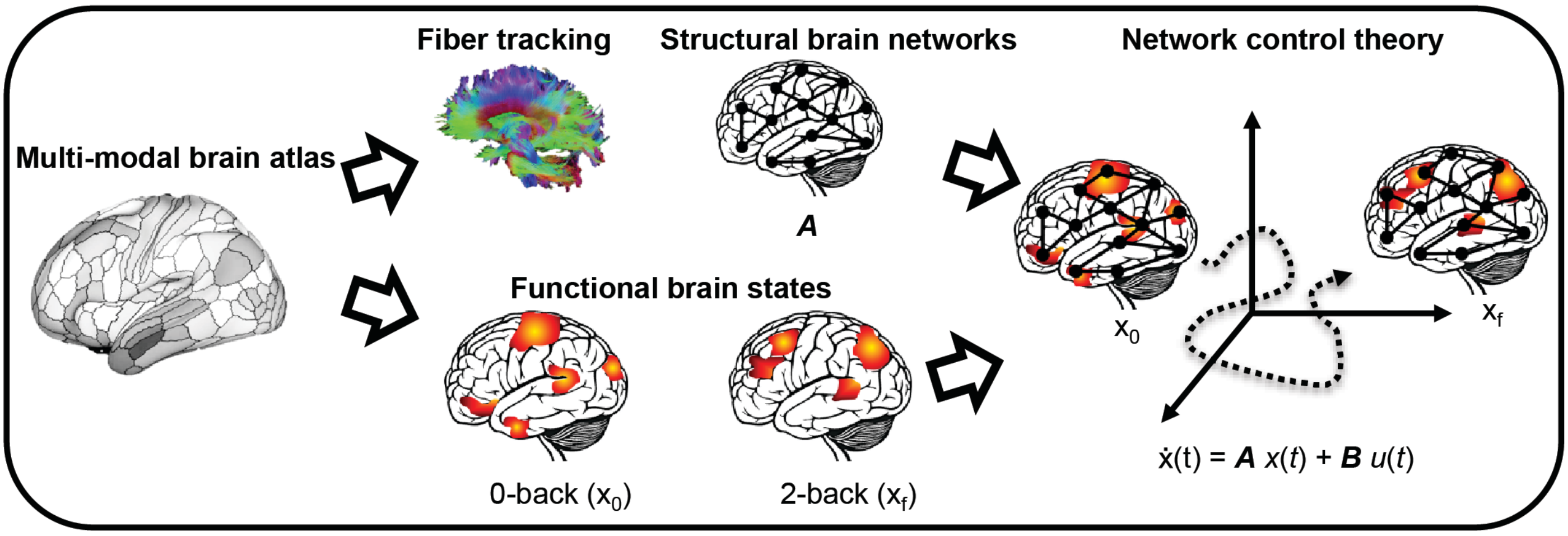

A summary of the methods to assess brain dynamics using network control theory. We used a multimodal atlas, and applied it to both diffusion tensor imaging data to obtain a structural connectome, and to functional magnetic resonance imaging data to obtain activation patterns during a 0-back motor and attention control task and during a 2-back working memory task. Finally, we use network control theory to explain transitions between 0-back and 2-back states based on the underlying structural connectome.

# 11. Supplementary tables

## 11.1. Table S1: Characteristics of the healthy control and patient samples

| | Healthy controls (n = 178) | Matched controls (n = 80) | Schizophrenia patients (n = 24) | t or χ² value | *P* value |
|---|---|---|---|---|---|
| ***Demographic information*** | | | | | |
| Age (year) | 33.05 ± 10.98 | 35.49 ± 10.55 | 32.25 ± 10.33 | 1.32 | 0.188 |
| Sex (male / female) | 93 / 85 | 46 / 34 | 18 / 6 | 2.39 | 0.122 |
| Years of education | 13.66 ± 2.41 | 13.65 ± 2.73 | 11.68 ± 1.45 | 2.72 | 0.008 |
| **Psychological assessments** | | | | | |
| MWTB | 30.74 ± 3.84 | 30.32 ± 4.83 | 29.13 ±3.27 | 1.11 | 0.272 |
| PANSS positive | n.a | n.a. | 12.50 ± 6.76 | - | - |
| PANSS negative | n.a | n.a | 15.17 ± 6.76 | - | - |
| BDI | n.a | n.a. | 12.42 ± 7.71 | - | - |
| Years of illness | n.a. | n.a. | 10.22 ± 9.32 | - | - |
| ***fMRI task performance*** | | | | | |
| Accuracy (%) | 80.10 ± 18.35 | 68.75 ± 19.33 | 65.54 ± 19.79 | 0.70 | 0.479 |
| Reaction time (ms) | 496.15 ± 279.50 | 589.21 ± 286.80 | 627.06 ± 306.99 | -0.6 | 0.553 |
| ***Imaging quality parameters*** | | | | | |
| fMRI: mean frame-wise displacement (mm) | 0.15 ± 0.06 | 0.18 ± 0.07 | 0.20 ± 0.09 | -1.11 | 0.270 |
| DTI: mean absolute root-mean-square displacement (mm) | 1.27 ± 0.74 | 1.37 ± 0.89 | 1.34 ± 0.59 | 0.18 | 0.860 |
| DTI: tSNR | 5.61 ± 0.45 | 5.52 ± 0.49 | 5.26 ± 0.49 | 2.21 | 0.028 |

Abbreviations: MWTB = Mehrfach Wortschatz Intelligenztest B, a German multiple-choice vocabulary intelligence test as a measure of premorbid IQ; PANSS = positive and negative symptom scale; BDI = Beck’s depression inventory; DTI = diffusion tensor imaging; tSNR temporal signal-to-noise.

## 11.2. Table S2: Sample characteristics of the pharmacological study

| | Healthy control (n = 16) | | t value | *P* value |
|---|---|---|---|---|
| | **Placebo** | **Amisulpride** | | |
| ***Demographic information*** | | | | |
| Age (year) | 26.63 ± 5.34 | | | |
| Sex (male / female) | 8 / 8 | | | |
| ***fMRI task performance **** | | | | |
| Accuracy (%) | 89.32 ± 10.00 | 85.68 ± 11.66 | 1.27 | 0.223 |
| Reaction time (ms) | 347.42 ± 104.26 | 365.40 ± 142.54 | -0.582 | 0.569 |
| ***Head motion parameters*** | | | | |
| fMRI: Mean frame-wise displacement (mm) | 0.124 ± 0.03 | 0.126 ± 0.04 | -0.301 | 0.767 |

## 11.3. Table S3: Statistical details for the main findings and replication analyses

| Result | Glasser | | | | | | Gordon | | |
|---|---|---|---|---|---|---|---|---|---|
| | FA | | | count | | | FA | | |
| | *df* | *F or t* (stand beta) | *p-val* | *df* | *F or t* (stand beta) | *p-val* | *df* | *F or t* (stand beta) | *p-val* |
| ***General properties of control*** | | | | | | | | | |
| stability 0 back > 2 back | *1,173* | 66.80 | < 0.001 | *1,173* | 60.50 | <0.001 | *1,173* | 56.846 | <0.001 |
| T02 > T20 | *1,174* | 27.98 | <0.001 | *1,174* | 19.73 | <0.001 | *1,174* | 21.706 | <0.001 |
| stability 2-back -> accuracy | | -2.78 | 0.006 | | 2.70 | 0.008 | | 2.58 | 0.011 |
| | | *(0.274)* | | | *(0.252)* | | | *(-0.239)* | |
| stability 2-back -> RT | | -1.94 | 0.054 | | -1.94 | 0.054 | | -1.95 | 0.053 |
| | | *(-0.192)* | | | *(-0.182)* | | | *(-0.181)* | |
| ***Differential relation to D1 and D2 expression*** | | | | | | | | | |
| D1 -> stability 0 back | | 2.18 | 0.034 | | *2.212* | 0.031 | | 3.10 | 0.003 |
| | | *(0.184)* | | | *(0.190)* | | | *(0.270)* | |
| D1 -> stability 2 back | | 2.78 | 0.007 | | 2.978 | 0.004 | | 3.39 | 0.001 |
| | | *(0.242)* | | | *(0.270)* | | | *(0.307)* | |
| D2 -> stability 0 back | | *1.629* | 0.109 | | *1.963* | 0.055 | | *0.146* | *0.149* |
| | | *(0.153)* | | | *(0.185)* | | | *(0.147)* | |
| D2 -> stability 2 back | | *-0.095* | 0.924 | | *0071* | 0.944 | | *0.174* | *0.862* |
| | | *(-0.010)* | | | *(0.008)* | | | *(0.019)* | |
| D2 -> T02 | | -2.14 | 0.037 | | -2.33 | 0.023 | | -2.07 | 0.043 |
| | | *(-0.076)* | | | *(-0.09)* | | | *(-0.07)* | |
| D2 -> T20 | | -1.87 | 0.068 | | -1.83 | 0.073 | | -2.16 | 0.036 |
| | | *(-0.134)* | | | *(-0.134)* | | | *(-0.152)* | |
| D1 -> T02 | | *-0.996* | 0.324 | | *-1.066* | 0.291 | | *-0.832* | *0.414* |
| | | *(-0.037)* | | | *(-0.045)* | | | *(-0.029)* | |
| D1 -> T20 | | *-0.817* | 0.418 | | *-0.650* | 0.519 | | *-0.376* | *0.709* |
| | | *(-0.06)* | | | *(-0.049)* | | | *(-0.028)* | |
| ***Drug effect on transition energy*** | | | | | | | | | |
| Pla vs. Ami | *1,10* | 7.272 | 0.022 | *1,10* | 4.954 | 0.05 | *1,10* | 8.839 | 0.014 |
| Pla vs. Ris | *1,10* | 3.49 | 0.091 | *1,10* | 3.797 | 0.08 | *1,10* | 3.014 | 0.113 |
| ***Drug effect on stability*** | | | | | | | | | |
| Pla vs. Ami | *1,8* | 0.715 | 0.422 | *1,8* | 0.698 | 0.428 | *1,8* | 0.013 | 0.913 |
| Pla vs. Ris | *1,8* | *1.057* | *0.334* | *1,8* | 2.941 | 0.125 | *1,8* | 0.358 | 0.566 |
| ***Schizophrenia*** | | | | | | | | | |
| stability 2-back: HC vs. SZ | *1,98* | 6.436 | 0.013 | *1,98* | 6.552 | 0.012 | *1,98* | 4.951 | 0.028 |
| stability 0-back: HC vs. SZ | *1,98* | 0.041 | 0.840 | *1,98* | 0.105 | 0.746 | *1,98* | 0.327 | 0.569 |
| T02: HC vs. SZ | *1,98* | 5.238 | 0.024 | *1,98* | 6.414 | 0.013 | *1,98* | 5.070 | 0.027 |
| T20: HC vs. SZ | *1,98* | 0.620 | 0.433 | *1,98* | 0.534 | 0.467 | *1,98* | 0.275 | 0.601 |

Abbreviations: FA = structural edge weight defined as mean fraction anisotropy of a track connecting two regions; Count = structural edge weight defined as track count connecting two regions; Gordon = resting-state defined atlas with 333 regions; T02 = control energy for the transition 0 to 2-back; T20 = control energy for the transition 2 to 0-back; -> = predicts in a regression model; Q = modularity estimate; Pla = Placebo; Ami = Amisulpride; Ris = Risperidone; HC = healthy control; SZ = schizophrenia patients.

### 11.4. Table S4: List of brain regions included in the extended Glasser parcellation

| # | Brain Region Label |
|---|---|
| 1 | Glasser_L_V1 |
| 2 | Glasser_L_MST |
| 3 | Glasser_L_V6 |
| 4 | Glasser_L_V2 |
| 5 | Glasser_L_V3 |
| 6 | Glasser_L_V4 |
| 7 | Glasser_L_V8 |
| 8 | Glasser_L_4 |
| 9 | Glasser_L_3b |
| 10 | Glasser_L_FEF |
| 11 | Glasser_L_PEF |
| 12 | Glasser_L_55b |
| 13 | Glasser_L_V3A |
| 14 | Glasser_L_RSC |
| 15 | Glasser_L_POS2 |
| 16 | Glasser_L_V7 |
| 17 | Glasser_L_IPS1 |
| 18 | Glasser_L_FFC |
| 19 | Glasser_L_V3B |
| 20 | Glasser_L_LO1 |
| 21 | Glasser_L_LO2 |
| 22 | Glasser_L_PIT |
| 23 | Glasser_L_MT |
| 24 | Glasser_L_A1 |
| 25 | Glasser_L_PSL |
| 26 | Glasser_L_SFL |
| 27 | Glasser_L_PCV |
| 28 | Glasser_L_STV |
| 29 | Glasser_L_7Pm |
| 30 | Glasser_L_7m |
| 31 | Glasser_L_POS1 |
| 32 | Glasser_L_23d |
| 33 | Glasser_L_v23ab |
| 34 | Glasser_L_d23ab |
| 35 | Glasser_L_31pv |
| 36 | Glasser_L_5m |
| 37 | Glasser_L_5mv |
| 38 | Glasser_L_23c |
| 39 | Glasser_L_5L |
| 40 | Glasser_L_24dd |
| 41 | Glasser_L_24dv |
| 42 | Glasser_L_7AL |
| 43 | Glasser_L_SCEF |
| 44 | Glasser_L_6ma |
| 45 | Glasser_L_7Am |
| 46 | Glasser_L_7PL |
| 47 | Glasser_L_7PC |
| 48 | Glasser_L_LIPv |
| 49 | Glasser_L_VIP |
| 50 | Glasser_L_MIP |
| 51 | Glasser_L_1 |
| 52 | Glasser_L_2 |
| 53 | Glasser_L_3a |
| 54 | Glasser_L_6d |
| 55 | Glasser_L_6mp |
| 56 | Glasser_L_6v |
| 57 | Glasser_L_p24pr |
| 58 | Glasser_L_33pr |
| 59 | Glasser_L_a24pr |
| 60 | Glasser_L_p32pr |
| 61 | Glasser_L_a24 |
| 62 | Glasser_L_d32 |
| 63 | Glasser_L_8BM |
| 64 | Glasser_L_p32 |
| 65 | Glasser_L_10r |
| 66 | Glasser_L_47m |
| 67 | Glasser_L_8Av |
| 68 | Glasser_L_8Ad |
| 69 | Glasser_L_9m |
| 70 | Glasser_L_8BL |
| 71 | Glasser_L_9p |
| 72 | Glasser_L_10d |
| 73 | Glasser_L_8C |
| 74 | Glasser_L_44 |
| 75 | Glasser_L_45 |
| 76 | Glasser_L_47l |
| 77 | Glasser_L_a47r |
| 78 | Glasser_L_6r |
| 79 | Glasser_L_IFJa |
| 80 | Glasser_L_IFJp |
| 81 | Glasser_L_IFSp |
| 82 | Glasser_L_IFSa |
| 83 | Glasser_L_p9-46v |
| 84 | Glasser_L_46 |
| 85 | Glasser_L_a9-46v |
| 86 | Glasser_L_9-46d |
| 87 | Glasser_L_9a |
| 88 | Glasser_L_10v |
| 89 | Glasser_L_a10p |
| 90 | Glasser_L_10pp |
| 91 | Glasser_L_11l |
| 92 | Glasser_L_13l |
| 93 | Glasser_L_OFC |
| 94 | Glasser_L_47s |
| 95 | Glasser_L_LIPd |
| 96 | Glasser_L_6a |
| 97 | Glasser_L_i6-8 |
| 98 | Glasser_L_s6-8 |
| 99 | Glasser_L_43 |
| 100 | Glasser_L_OP4 |
| 101 | Glasser_L_OP1 |
| 102 | Glasser_L_OP2-3 |
| 103 | Glasser_L_52 |
| 104 | Glasser_L_RI |
| 105 | Glasser_L_PFcm |
| 106 | Glasser_L_PoI2 |
| 107 | Glasser_L_TA2 |
| 108 | Glasser_L_FOP4 |
| 109 | Glasser_L_MI |
| 110 | Glasser_L_Pir |
| 111 | Glasser_L_AVI |
| 112 | Glasser_L_AAIC |
| 113 | Glasser_L_FOP1 |
| 114 | Glasser_L_FOP3 |
| 115 | Glasser_L_FOP2 |
| 116 | Glasser_L_PFt |
| 117 | Glasser_L_AIP |
| 118 | Glasser_L_EC |
| 119 | Glasser_L_PreS |
| 120 | Glasser_L_H |
| 121 | Glasser_L_ProS |
| 122 | Glasser_L_PeEc |
| 123 | Glasser_L_STGa |
| 124 | Glasser_L_PBelt |
| 125 | Glasser_L_A5 |
| 126 | Glasser_L_PHA1 |
| 127 | Glasser_L_PHA3 |
| 128 | Glasser_L_STSda |
| 129 | Glasser_L_STSdp |
| 130 | Glasser_L_STSvp |
| 131 | Glasser_L_TGd |
| 132 | Glasser_L_TE1a |
| 133 | Glasser_L_TE1p |
| 134 | Glasser_L_TE2a |
| 135 | Glasser_L_TF |
| 136 | Glasser_L_TE2p |
| 137 | Glasser_L_PHT |
| 138 | Glasser_L_PH |
| 139 | Glasser_L_TPOJ1 |
| 140 | Glasser_L_TPOJ2 |
| 141 | Glasser_L_TPOJ3 |
| 142 | Glasser_L_DVT |
| 143 | Glasser_L_PGp |
| 144 | Glasser_L_IP2 |
| 145 | Glasser_L_IP1 |
| 146 | Glasser_L_IP0 |
| 147 | Glasser_L_PFop |
| 148 | Glasser_L_PF |
| 149 | Glasser_L_PFm |
| 150 | Glasser_L_PGi |
| 151 | Glasser_L_PGs |
| 152 | Glasser_L_V6A |
| 153 | Glasser_L_VMV1 |
| 154 | Glasser_L_VMV3 |
| 155 | Glasser_L_PHA2 |
| 156 | Glasser_L_V4t |
| 157 | Glasser_L_FST |
| 158 | Glasser_L_V3CD |
| 159 | Glasser_L_LO3 |
| 160 | Glasser_L_VMV2 |
| 161 | Glasser_L_31pd |
| 162 | Glasser_L_31a |
| 163 | Glasser_L_VVC |
| 164 | Glasser_L_25 |
| 165 | Glasser_L_s32 |
| 166 | Glasser_L_pOFC |
| 167 | Glasser_L_PoI1 |
| 168 | Glasser_L_Ig |
| 169 | Glasser_L_FOP5 |
| 170 | Glasser_L_p10p |
| 171 | Glasser_L_p47r |
| 172 | Glasser_L_TGv |
| 173 | Glasser_L_MBelt |
| 174 | Glasser_L_LBelt |
| 175 | Glasser_L_A4 |
| 176 | Glasser_L_STSva |
| 177 | Glasser_L_TE1m |
| 178 | Glasser_L_PI |
| 179 | Glasser_L_a32pr |
| 180 | Glasser_L_p24 |
| 181 | Glasser_R_V1 |
| 182 | Glasser_R_MST |
| 183 | Glasser_R_V6 |
| 184 | Glasser_R_V2 |
| 185 | Glasser_R_V3 |
| 186 | Glasser_R_V4 |
| 187 | Glasser_R_V8 |
| 188 | Glasser_R_4 |
| 189 | Glasser_R_3b |
| 190 | Glasser_R_FEF |
| 191 | Glasser_R_PEF |
| 192 | Glasser_R_55b |
| 193 | Glasser_R_V3A |
| 194 | Glasser_R_RSC |
| 195 | Glasser_R_POS2 |
| 196 | Glasser_R_V7 |
| 197 | Glasser_R_IPS1 |
| 198 | Glasser_R_FFC |
| 199 | Glasser_R_V3B |
| 200 | Glasser_R_LO1 |
| 201 | Glasser_R_LO2 |
| 202 | Glasser_R_PIT |
| 203 | Glasser_R_MT |
| 204 | Glasser_R_A1 |
| 205 | Glasser_R_PSL |
| 206 | Glasser_R_SFL |
| 207 | Glasser_R_PCV |
| 208 | Glasser_R_STV |
| 209 | Glasser_R_7Pm |
| 210 | Glasser_R_7m |
| 211 | Glasser_R_POS1 |
| 212 | Glasser_R_23d |
| 213 | Glasser_R_v23ab |
| 214 | Glasser_R_d23ab |
| 215 | Glasser_R_31pv |
| 216 | Glasser_R_5m |
| 217 | Glasser_R_5mv |
| 218 | Glasser_R_23c |
| 219 | Glasser_R_5L |
| 220 | Glasser_R_24dd |
| 221 | Glasser_R_24dv |
| 222 | Glasser_R_7AL |
| 223 | Glasser_R_SCEF |
| 224 | Glasser_R_6ma |
| 225 | Glasser_R_7Am |
| 226 | Glasser_R_7PL |
| 227 | Glasser_R_7PC |
| 228 | Glasser_R_LIPv |
| 229 | Glasser_R_VIP |
| 230 | Glasser_R_MIP |
| 231 | Glasser_R_1 |
| 232 | Glasser_R_2 |
| 233 | Glasser_R_3a |
| 234 | Glasser_R_6d |
| 235 | Glasser_R_6mp |
| 236 | Glasser_R_6v |
| 237 | Glasser_R_p24pr |
| 238 | Glasser_R_33pr |
| 239 | Glasser_R_a24pr |
| 240 | Glasser_R_p32pr |
| 241 | Glasser_R_a24 |
| 242 | Glasser_R_d32 |
| 243 | Glasser_R_8BM |
| 244 | Glasser_R_p32 |
| 245 | Glasser_R_10r |
| 246 | Glasser_R_47m |
| 247 | Glasser_R_8Av |
| 248 | Glasser_R_8Ad |
| 249 | Glasser_R_9m |
| 250 | Glasser_R_8BL |
| 251 | Glasser_R_9p |
| 252 | Glasser_R_10d |
| 253 | Glasser_R_8C |
| 254 | Glasser_R_44 |
| 255 | Glasser_R_45 |
| 256 | Glasser_R_47l |
| 257 | Glasser_R_a47r |
| 258 | Glasser_R_6r |
| 259 | Glasser_R_IFJa |
| 260 | Glasser_R_IFJp |
| 261 | Glasser_R_IFSp |
| 262 | Glasser_R_IFSa |
| 263 | Glasser_R_p9-46v |
| 264 | Glasser_R_46 |
| 265 | Glasser_R_a9-46v |
| 266 | Glasser_R_9-46d |
| 267 | Glasser_R_9a |
| 268 | Glasser_R_10v |
| 269 | Glasser_R_a10p |
| 270 | Glasser_R_10pp |
| 271 | Glasser_R_11l |
| 272 | Glasser_R_13l |
| 273 | Glasser_R_OFC |
| 274 | Glasser_R_47s |
| 275 | Glasser_R_LIPd |
| 276 | Glasser_R_6a |
| 277 | Glasser_R_i6-8 |
| 278 | Glasser_R_s6-8 |
| 279 | Glasser_R_43 |
| 280 | Glasser_R_OP4 |
| 281 | Glasser_R_OP1 |
| 282 | Glasser_R_OP2-3 |
| 283 | Glasser_R_52 |
| 284 | Glasser_R_RI |
| 285 | Glasser_R_PFcm |
| 286 | Glasser_R_PoI2 |
| 287 | Glasser_R_TA2 |
| 288 | Glasser_R_FOP4 |
| 289 | Glasser_R_MI |
| 290 | Glasser_R_Pir |
| 291 | Glasser_R_AVI |
| 292 | Glasser_R_AAIC |
| 293 | Glasser_R_FOP1 |
| 294 | Glasser_R_FOP3 |
| 295 | Glasser_R_FOP2 |
| 296 | Glasser_R_PFt |
| 297 | Glasser_R_AIP |
| 298 | Glasser_R_EC |
| 299 | Glasser_R_PreS |
| 300 | Glasser_R_H |
| 301 | Glasser_R_ProS |
| 302 | Glasser_R_PeEc |
| 303 | Glasser_R_STGa |
| 304 | Glasser_R_PBelt |
| 305 | Glasser_R_A5 |
| 306 | Glasser_R_PHA1 |
| 307 | Glasser_R_PHA3 |
| 308 | Glasser_R_STSda |
| 309 | Glasser_R_STSdp |
| 310 | Glasser_R_STSvp |
| 311 | Glasser_R_TGd |
| 312 | Glasser_R_TE1a |
| 313 | Glasser_R_TE1p |
| 314 | Glasser_R_TE2a |
| 315 | Glasser_R_TF |
| 316 | Glasser_R_TE2p |
| 317 | Glasser_R_PHT |
| 318 | Glasser_R_PH |
| 319 | Glasser_R_TPOJ1 |
| 320 | Glasser_R_TPOJ2 |
| 321 | Glasser_R_TPOJ3 |
| 322 | Glasser_R_DVT |
| 323 | Glasser_R_PGp |
| 324 | Glasser_R_IP2 |
| 325 | Glasser_R_IP1 |
| 326 | Glasser_R_IP0 |
| 327 | Glasser_R_PFop |
| 328 | Glasser_R_PF |
| 329 | Glasser_R_PFm |
| 330 | Glasser_R_PGi |
| 331 | Glasser_R_PGs |
| 332 | Glasser_R_V6A |
| 333 | Glasser_R_VMV1 |
| 334 | Glasser_R_VMV3 |
| 335 | Glasser_R_PHA2 |
| 336 | Glasser_R_V4t |
| 337 | Glasser_R_FST |
| 338 | Glasser_R_V3CD |
| 339 | Glasser_R_LO3 |
| 340 | Glasser_R_VMV2 |
| 341 | Glasser_R_31pd |
| 342 | Glasser_R_31a |
| 343 | Glasser_R_VVC |
| 344 | Glasser_R_25 |
| 345 | Glasser_R_s32 |
| 346 | Glasser_R_pOFC |
| 347 | Glasser_R_PoI1 |
| 348 | Glasser_R_Ig |
| 349 | Glasser_R_FOP5 |
| 350 | Glasser_R_p10p |
| 351 | Glasser_R_p47r |
| 352 | Glasser_R_TGv |
| 353 | Glasser_R_MBelt |
| 354 | Glasser_R_LBelt |
| 355 | Glasser_R_A4 |
| 356 | Glasser_R_STSva |
| 357 | Glasser_R_TE1m |
| 358 | Glasser_R_PI |
| 359 | Glasser_R_a32pr |
| 360 | Glasser_R_p24 |
| 504 | HO_Left_Thalamus |
| 505 | HO_Left_Caudate |
| 506 | HO_Left_Putamen |
| 507 | HO_Left_Pallidum |
| 509 | HO_Left_Hippocampus |
| 510 | HO_Left_Amygdala |
| 511 | HO_Left_Accumbens |
| 515 | HO_Right_Thalamus |
| 516 | HO_Right_Caudate |
| 517 | HO_Right_Putamen |
| 518 | HO_Right_Pallidum |
| 519 | HO_Right_Hippocampus |
| 520 | HO_Right_Amygdala |
| 521 | HO_Right_Accumbens |

## 12. Supplementary references

1. Tombaugh TN (2004) Trail Making Test A and B: normative data stratified by age and education. *Archives of clinical neuropsychology* 19(2):203-214.
2. Lehrl S, Triebig G, & Fischer B (1995) Multiple choice vocabulary test MWT as a valid and short test to estimate premorbid intelligence. *Acta Neurologica Scandinavica* 91(5):335-345.
3. Kay SR, Fiszbein A, & Opler LA (1987) The positive and negative syndrome scale (PANSS) for schizophrenia. *Schizophrenia bulletin* 13(2):261-276.
4. Beck AT, Ward CH, Mendelson M, Mock J, & Erbaugh J (1961) An inventory for measuring depression. *Arch Gen Psychiatry* 4:561-571.
5. Braun U*, et al.* (2015) Dynamic reconfiguration of frontal brain networks during executive cognition in humans. *Proc Natl Acad Sci U S A* 112(37):11678-11683.
6. Plichta MM*, et al.* (2012) Test-retest reliability of evoked BOLD signals from a cognitive-emotive fMRI test battery. *Neuroimage* 60(3):1746-1758.
7. Callicott JH*, et al.* (1999) Physiological characteristics of capacity constraints in working memory as revealed by functional MRI. *Cerebral Cortex* 9(1):20-26.
8. Glasser MF*, et al.* (2016) A multi-modal parcellation of human cerebral cortex. *Nature* 536(7615):171-178.
9. Smith SM*, et al.* (2004) Advances in functional and structural MR image analysis and implementation as FSL. *Neuroimage* 23 Suppl 1:S208-219.
10. Smith SM (2002) Fast robust automated brain extraction. *Hum Brain Mapp* 17(3):143-155.
11. Yeh FC, Verstynen TD, Wang Y, Fernandez-Miranda JC, & Tseng WY (2013) Deterministic diffusion fiber tracking improved by quantitative anisotropy. *PLoS One* 8(11):e80713.
12. Baum GL*, et al.* (2018) The impact of in-scanner head motion on structural connectivity derived from diffusion MRI. *Neuroimage* 173:275-286.
13. Chung MH*, et al.* (2018) Individual differences in rate of acquiring stable neural representations of tasks in fMRI. *PLoS One* 13(11):e0207352.
14. Braun U*, et al.* (2016) Dynamic brain network reconfiguration as a potential schizophrenia genetic risk mechanism modulated by NMDA receptor function. *Proc Natl Acad Sci U S A* 113(44):12568-12573.
15. Betzel RF, Gu S, Medaglia JD, Pasqualetti F, & Bassett DS (2016) Optimally controlling the human connectome: the role of network topology. *Sci Rep* 6:30770.
16. Gu S*, et al.* (2017) Optimal trajectories of brain state transitions. *Neuroimage* 148:305-317.
17. Kim JZ*, et al.* (2018) Role of Graph Architecture in Controlling Dynamical Networks with Applications to Neural Systems. *Nat Phys* 14:91-98.
18. Cornblath EJ*, et al.* (2018) Context-dependent architecture of brain state dynamics is explained by white matter connectivity and theories of network control. *arXiv preprint arXiv:1809.02849*.
19. Howie B, Fuchsberger C, Stephens M, Marchini J, & Abecasis GR (2012) Fast and accurate genotype imputation in genome-wide association studies through pre-phasing. *Nat Genet* 44(8):955-959.
20. Howie BN, Donnelly P, & Marchini J (2009) A flexible and accurate genotype imputation method for the next generation of genome-wide association studies. *PLoS genetics* 5(6):e1000529.
21. Delaneau O, Zagury JF, & Marchini J (2013) Improved whole-chromosome phasing for disease and population genetic studies. *Nature methods* 10(1):5-6.
22. Fazio L*, et al.* (2018) Transcriptomic context of DRD1 is associated with prefrontal activity and behavior during working memory. *Proc Natl Acad Sci U S A* 115(21):5582-5587.
23. Pergola G*, et al.* (2017) DRD2 co-expression network and a related polygenic index predict imaging, behavioral and clinical phenotypes linked to schizophrenia. *Transl Psychiatry* 7(1):e1006.
24. Selvaggi P*, et al.* (2019) Genetic Variation of a DRD2 Co-expression Network is Associated with Changes in Prefrontal Function After D2 Receptors Stimulation. *Cereb Cortex* 29(3):1162-1173.
25. Fazio L*, et al.* (2018) Transcriptomic context of DRD1 is associated with prefrontal activity and behavior during working memory. *Proceedings of the National Academy of Sciences of the United States of America*.
26. Selvaggi P*, et al.* (2018) Genetic Variation of a DRD2 Co-expression Network is Associated with Changes in Prefrontal Function After D2 Receptors Stimulation. *Cerebral cortex*.

27. Pergola G, *et al.* (2018) Prefrontal co-expression of schizophrenia risk genes is associated with treatment response in patients. *bioRxiv*.
28. Zhang B & Horvath S (2005) A general framework for weighted gene co-expression network analysis. *Statistical applications in genetics and molecular biology* 4:Article17.
29. Colantuoni C, *et al.* (2011) Temporal dynamics and genetic control of transcription in the human prefrontal cortex. *Nature* 478(7370):519-523.
30. Pergola G, *et al.* (2016) Combined effect of genetic variants in the GluN2B coding gene (GRIN2B) on prefrontal function during working memory performance. *Psychological medicine* 46(6):1135-1150.
31. Gordon EM, *et al.* (2016) Generation and Evaluation of a Cortical Area Parcellation from Resting-State Correlations. *Cereb Cortex* 26(1):288-303.